\documentclass[12pt]{amsart}

\usepackage{times,alltt,fullpage,graphics,amsmath, amssymb, version}
\usepackage{amssymb,verbatim,boxedminipage}

\usepackage[top=3cm,bottom=3cm,left=3cm,right=3cm, textwidth = 6.5in]{geometry}

\newcommand{\noks}{N_{0,k,s}}
\newcommand{\naks}{N_{1,k,s}}
\newcommand{\nbks}{N_{2,k,s}}
\newcommand{\noksp}{N_{0,k,s+1}}
\newcommand{\bP}{{\mathbb P}}

\newcommand{\Pqs}{{\mathbb P}_{q}^s}
\newcommand{\Pqks}{{{\mathbb P}_{q^k}^s}}
\newcommand{\piqks}{{\pi}_{q^k ,s}}

\newcommand{\Npqks}{N_{k,s}^{\prime}}

\newcommand{\F}{{\mathbb F}}

\newcommand{\Fq}{{{\F}_q}}

\newcommand{\Fqk}{{\F}_{q^k}}

\newcommand{\cH}{{\mathcal H}}
\newcommand{\cHs}{{\cH}_s}
\newcommand{\cA}{{\mathcal A}}
\newcommand{\cAs}{{\cA}_s}

\def\qed{{\hfill $\Box$ \medbreak}}

\newtheorem{defi}{Definition}[section]
\newtheorem{thm}[defi]{Theorem}
\newtheorem{lem}[defi]{Lemma}

\def\F{\mathbb{F}}

\title[Diagonal and Hermitian surfaces]{A note on diagonal and Hermitian surfaces}
\author{\textsc{Ian Blake}}
\address{Department of Electrical and Computer Engineering, University of British Columbia, Vancouver, BC, V6T 1Z4}
\email{ifblake@ece.ubc.ca}
\author{V. Kumar Murty}
\address{Department of Mathematics. University of Toronto, Toronto, ON M5S 2E4}
\email{murty@math.toronto.edu}
\author{\textsc{Hamid Usefi}}
\address{Department of Mathematics, Memorial University, St. John's, NL, A1C 5S7}
\email{usefi@mun.ca}
\thanks{ The research of the first and second author are supported by NSERC. The research of the third author is supported by NSERC and the Research \& Development Corporation of Newfoundland and Labrador}

\date{\today}

\begin{document}

\begin{abstract}
Aspects of the properties, enumeration and construction
of points on diagonal and Hermitian surfaces 
have been considered extensively in the literature and are 
further considered here. The zeta function of diagonal surfaces
is given as a direct result of the work of Wolfmann. 
Recursive construction techniques for the set of 
rational points of Hermitian surfaces are of interest. 
The relationship of these techniques here to 
the construction of codes on surfaces is briefly noted.
\end{abstract}

\subjclass[2000]{}

\maketitle

\section{Introduction}

Th enumeration of points on hypersurfaces over finite fields has been 
a problem of interest in the study of algebraic geometry, culminating
in the celebrated conjectures and theorems of Weil and Deligne.
The case of surfaces defined by diagonal equations has been
of particular interest and the work of Wolfmann \cite{wolf} 
gives expressions for these for the cases of interest in this work.
These expressions are examined further here with a view to understanding
their properties better.

The next Section \ref{sec:diag} considers the case of surfaces from
diagonal equations and, in particular, the important results 
of Wolfmann \cite{wolf} on the enumeration of diagonal
surfaces. The properties of functions introduced by Wolfmann
are explored.  Section \ref{sec:zeta} gives a brief review of aspects of 
algebraic curves and hypersurfaces over a finite field and their zeta functions. 
In particular, the zeta function of diagonal surfaces is seen to follow as a 
direct consequence of this work.

One of the motivations of this work was to apply the recursive construction 
of the points on Hermitian surfaces and their properties to derive results on 
the minimum distance of codes obtained from Hermitian surfaces,
a problem that has attracted considerable attention. While this approach has 
so far been unsuccessful, the problem is discussed in
Section \ref{sec:coding}.
\vspace{.1in}

\section{Enumeration of diagonal surfaces}
\label{sec:diag}

While much of the remainder of the paper will consider projective surfaces
over the finite field $\Fq$, this section will consider the affine case
and the results of Wolfmann \cite{wolf}. So far as possible the notation 
of Wolfmann will be used although there will be important
differences to accommodate our interests in the projective 
cases in later sections.

Denote by $\noks$ the number of solutions over $\Fqk$ of the
equation
\begin{equation}
\label{eq:diag}
x_1^d + x_2^d + \cdots + x_s^d =0
\end{equation}
where $q=p^{2r}$ and $d \mid p^r +1$. The parameter $d$ will be 
fixed throughout. Define the function $B(d,s)$ as:

\begin{equation}
\label{eq:bds}
B(d,s) = \frac{1}{d} \left( (d-1)^s + (-1)^s (d-1) \right) .
\end{equation}

The following result of Wolfmann is central to this work.

\begin{thm}[\cite{wolf}, Corollary 4]
\label{thm:wolf}
Let $p$ be a prime number, $q=p^{2r}$, and $\Fqk$ the finite field of order $q^k$. 
The number of solutions to the equation
\[
x_1^d + x_2^d + \cdots + x_s^d = b , \quad b \in \Fqk
\]
for $d \mid p^r +1$, $nd = q^k -1$ and $s \geq 2$ is as follows:
\begin{enumerate}
\item If $b=0$ we have $\noks := q^{k(s-1)}  +\eta^{s}  q^{k(s/2 -1)}(q^k-1)B(d,s)$.
\item If $b\neq 0$ and $b^n = 1$ then  $\naks :=q^{k(s -1)} +\eta^{s+1}  q^{k(s/2 -1)} [ (d-1)^{s} q^{k/2} -(q^{k/2}+\eta)B(d, s)]$.
\item If $b\neq 0$ and $b^n \neq 1$ then  $ \nbks :=q^{k (s -1)} +\eta^{s+1}  q^{k(s/2-1)} [ (-1)^{s}  q^{k/2} -(q^{k/2}+\eta)B(d, s)]$,
\end{enumerate}
where $\eta=(-1)^{k+1}$ and $B(d,s)$ is as in Equation (\ref{eq:bds}).
\end{thm}

Let $\alpha$ be a primitive root of $\Fqk$.
Notice that the equation $b^n = 1$ has $n$ solutions in 
$\Fqk$ if and only if  $b = \alpha^{jd}$ for some integer $j$.
Define the $n$-th roots of unity as $U_n = \{ \alpha^{jd}, 
j=0,1, \dots , (n-1) \}$ and notice that the equation $x^d = a$
has solutions if and only if  $a \in U_n$. If $\alpha^i$ is a solution of
this equation, the other solutions are $\alpha^{i + \ell n}$ for
$\ell = 0,1, \dots , (d-1)$.

Also notice that $-1 \in U_n$. This is argued as follows.
If $p$ is even (characteristic 2) then $-1=+1$ and is in $U_n$. If $p$
is odd then $n= (q^k -1)/d = (p^{2rk} -1)/d = (p^{rk} -1)(p^{rk} +1)/d$
and since by assumption $d \mid (p^r +1)$ it divides either the first or 
second factor. As both factors are even, so is $n$.
Thus if $a \in U_n$ then so is $-a$.

Many relationships among the quantities mentioned in the
above Theorem can be formulated. Three of these are noted below
as of sufficient interest to prove.

\begin{lem}
\label{lemma:1}
With the notation of Theorem (\ref{thm:wolf}), we have
\[
\noksp = \noks + (q^k -1) \naks .
\]
\end{lem}

\noindent
{\it Proof:}
The Lemma is a reflection of the fact that a solution to the
Equations (\ref{eq:diag}) with $s$ variables to one with $(s+1)$ variables 
can be developed in two ways, the first being by adding a zero for the
$(s+1)$-st variable to a solution to an $s$ variable one. The other considers
the situation where a solution ${\bf y} = (y_1 , y_2 , \dots , y_s )$ satisfies
\[
y_1^d + y_2^d + \cdots + y_s^d = b \in U_n .
\]
The equation $x^d = b \in U_n$ has $d$ solutions which can be added
as the $(s+1)$-st coordinate to the solution ${\bf y}$. There are such $n$ such values of $b$
giving rise to $nd$ such solutions and the second term of the Lemma 
as $nd = (q^k -1)$.

The equations of the Lemma can be expressed:
\[
\begin{array}{rl}
\noksp & = q^{k(s)}  +\eta^{s+1}  q^{k((s+1)/2 -1)}(q^k-1)B(d,s+1) \\
       & = q^{k(s-1)}  +\eta^{s}  q^{k(s/2 -1)}(q^k-1)B(d,s) + \\
       & (q^k -1)\left( q^{k(s -1)} +\eta^{s+1}  q^{k(s/2 -1)} [ (d-1)^{s} q^{k/2} -(q^{k/2}+\eta)B(d, s)] \right)
\end{array}
\]
Verification of this equation is straightforward.
\qed

\begin{lem}
\label{lemma:2}
With the notation of Theorem (\ref{thm:wolf}), we have
$q^{ks} = \noks + n \naks + (q^k -1-n) \nbks$.
\end{lem}

\noindent
{\it Proof:}
The Lemma reflects the fact that as the variables $x_i ,i=1,2, \dots , s$
range over all values of $\Fqk$, the sum $\sum_i x_i^d$ takes on values either
$0, ~ b \in U_n$ or $b \notin U_n$.
The direct proof, omitted here, involves verifying the equation:
\[
\begin{array}{rl}
q^{ks} = & q^{k(s-1)}  +\eta^{s}  q^{k(s/2 -1)}(q^k-1)B(d,s) \\
 & + n \cdot \big( q^{k(s -1)} +\eta^{s+1}  q^{k(s/2 -1)} [ (d-1)^{s} q^{k/2} -(q^{k/2}+\eta)B(d, s)] \big) \\
 & + (q^k - n -1) \cdot  \big( q^{k (s -1)} +\eta^{s+1}  q^{k(s/2-1)} [ (-1)^{s}  q^{k/2} -(q^{k/2}+\eta)B(d, s)] \big).
\end{array}
\]
\qed

\begin{lem}
\label{lemma:3}
Let $i$ be an integer in the range $1\leq i\leq s-1$. Then, with the notation of Theorem (\ref{thm:wolf}), we have
\[
N_{0,k,s} = N_{0,k,i} N_{0,k,s-i} + n N_{1,k,i} N_{1,k,s-i} + (q^k -1-n) N_{2,k,i} N_{2,k,s-i} .
\]
\end{lem}

\noindent
{\it Proof:}
The relation enumerates  the solutions to the equation 
\[
x_1^d + x_2^d + \cdots + x_i^d = a.
\]
If $a=0$, any such solution can be combined with a solution to $x_{i+1}^d + \cdots + x_s^d =0$
to yield the first term. 
Since if $a \in U_n$ then so is $-a$ and there are $n$ such values which 
gives the second term and similarly for the third term.

The formal verification involves substituting the expressions from 
Theorem (\ref{thm:wolf}) and is omitted.

\qed

Finally it is interesting to note that if, for a given value of $d$, the terms $N_{0,k,s}$
are known for all $k$ and $s$, then by applying Lemmas 2.2 and 2.3, they determine 
all other terms, $N_{1,k,s}$ and $N_{2,k,s}$. Thus the zero solutions determine uniquely 
all nonzero solutions.

\section{The zeta function of a diagonal surface}
\label{sec:zeta}

For the remainder of the work the projective case will be of interest and
the affine results of the previous section will be translated to the projective case.
The projective space of dimension $s$ over $\Fqk$,  denoted by $\Pqks$, is the set of $(s+1)$-tuples,
or points, over the finite field $\Fqk$ with scalar multiples identified. Thus
\begin{equation}
\label{eqn:proj}
\Pqks = \{ (a_0 , a_1 , \dots , a_s ) , ~ a_i \in \Fqk \} , \quad \mid \Pqks  \mid = (q^{k(s+1)} -1)/(q^k -1) = \piqks .
\end{equation}
Where needed, the representative 
of an equivalence class will have the first nonzero element in 
the $(s+1)$ tuple is unity. When $q^k$ is understood, we write ${\pi}_{q^k ,s}  = \pi_s$.

For the homogeneous polynomial $f(x_0 , x_1 , \dots , x_s )$ over $\Fqk$ define 
the projective variety 
\[
X_f ( \Fqk ) = \{ ( x_0 , x_1 , \dots , x_s ) \in \Pqks\mid ~ f(x_0 , x_1 , \dots , x_s ) =0 \}.
\]
The variety is referred to as a hypersurface when defined by a single polynomial as above.
It is also referred to as the zero set or {\it algebraic set} of $f$.
More general varieties over $\Fqk$ defined by sets of polynomials, will simply be denoted $X (\Fqk )$.
The hypersurface is called {\it nondegenerate} (\cite{litt}) if it is not contained
in a linear subspace of $\Pqks$, {\it smooth} if it contains no singularities 
(points where the function derivatives vanish simultaneously)
and irreducible if it is not the union of smaller algebraic sets. 
Many of the results considered have parallels in the associated affine case.

If the homogeneous polynomial $f$ with $(s+1)$ variables is of degree $d$ the variety 
is referred to as having dimension $s$ and degree $d$. The polynomials
of interest here, as in the previous section, are the diagonal ones:
\begin{equation}
\label{eqn:diag1}
x_0^d + x_1^d + \cdots + x_s^d = f(x_0, x_1 , \dots , x_s )
\end{equation}
The corresponding surfaces will be referred to as {\it diagonal}. 
The base field will be taken as $\Fq , q=p^{2r}$ for some prime $p$. 
When $d = p^r +1$ in the above equation, the variety over $\Fq$ will be 
referred to as {\it Hermitian}. When viewing this diagonal equation
over extensions of $\Fq$ it would cease to be Hermitian. 

The cardinality of projective varieties has been of great interest in the literature.
The case of affine varieties has been considered in the previous section. To translate
them to the projective case, denote the number of projective points of the variety 
defined by the diagonal polynomial over $\Fqk$ of Equation (\ref{eqn:diag1}) as $\Npqks$
and note that 

\begin{equation}
\label{eq:enum}
\Npqks = \frac{(\noksp -1)}{(q^k -1 )} = \mid X_f ( \Fqk ) \mid
\end{equation}
where primes will be used to denote the projective case and where $f$ is the diagonal polynomial
of Equation (\ref{eqn:diag1}).

Much effort has been spent on determining bounds on the number of points on curves and
surfaces.  An early result of Weil (following a conjecture of Hasse) for curves 
of genus $g$ is that
\[
\mid N - (1+q ) \mid \leq 2g q^{1/2},
\]
a result that was later improved by Serre to 
\[
\mid N - (1+q ) \mid \leq g \lfloor 2 q^{1/2} \rfloor .
\]

For the case of higher dimensional surfaces the generalized Weil-Deligne bound 
for a smooth nondegenerate hypersurface $X (\Fq )$, of dimension $s$ and degree 
$d$ is given by \cite{litt, sore2, wolf}:
\begin{equation}
\label{eq:size}
\mid X_{f} (\Fq )   - \pi_{q,s-1}  \mid \leq B(d,s+1) q^{(s-1)/2} , 
\end{equation}
where the function $B(d,s)$, given in Equation (\ref{eq:bds}), will be useful in much of the sequel.

The Tsfasman-Serre-Sorensen bound \cite{edou2,sore} for the same quantity is given by: 
\begin{equation}
\label{eq:TSS}
\mid X_{f} ( \Fq ) \mid \leq d q^{s-1} + \pi_{q, s-2} ,
\end{equation}
a relationship that can be used to establish the minimum distance of certain projective 
geometry codes, as discussed later. The bound of this last equation is achieved only if the 
zero set of the defining polynomial consists of a  union of $d$ hyperspaces with a common 
hyperspace of co-dimension 2 \cite{edou1,lach1}.

The zeta function of a curve or surface codifies the sizes of the variety
when viewed over extension fields. Specifically let $X ( \Fq )$ denote
a variety of dimension $s$ and degree $d$ defined over the finite field $\Fq$ and $X ( \Fqk )$
for the same variety viewed over the extension field $\Fqk$.
Let $N^{\prime}_{k,s}$ denote the number of projective solutions 
to this equation over ${\F}_{q^k}$ and define 
the zeta function for $X (\Fq )$ as
\[
Z_{s} (t)  = \exp \left( \sum_{k=1}^{\infty} \Npqks t^k /k \right),
\]
where the dependence on the variety is understood.

The following information on zeta functions is well known eg \cite{ghor, kobl, sore2, zarz}.
It is a rational function that can be expressed as
\[
Z_s (t) = \frac{P_1 (t) P_3 (t) \dots P_{2m-1} (t)}{P_0 (t) P_2 (t) \dots P_{2m} (t)}
\]
where the polynomials have integer coefficients and 
\[
P_i (t) = \prod_{j=1}^{b_i} (1- \omega_{ij} t), \quad 1 \leq i \leq 2m-1
\]
where $b_i$ is a Betti number, related to the topology of the variety,
 the $\omega_{ij}$ are algebraic integers and
$\mid \omega_{ij} \mid = q^{1/2}$ and $P_0 (t) = 1-t,  P_{2m} (t) = 1-q^m t$.
The polynomials have many properties and we refer the reader to \cite{ghor,kobl,zarz}.

In the case the variety is  smooth the zeta function can be refined to \cite{irel,kobl,sore1}:
\begin{thm}
\label{thm:weil}
Let $X_{f} (\Fq )$ be a smooth irreducible hypersurface over $\Fq$ in $\Pqs$
of degree $d$ and dimension $s$. Then
\[
Z_s (t) = \left\{
\begin{array}{ll}
\frac{P(t)}{\prod_{i=0}^{s-1} (1-q^i t)} , & s ~~ \text{even} \\
\frac{1}{P(t) \prod_{i=0}^{s-1} (1-q^i t)} , & s~~ \text{odd}
\end{array}
\right.
\]
where $P(t) = \prod_{i=1}^{B(d,s+1)} (1-\alpha_i t)$ and $B(d,s+1)$ is as in Equation (\ref{eq:bds}).
$P(t)$ has integer coefficients and $\mid \alpha_i \mid = q^{(s-1)/2}$.
\end{thm}

The results of Wolfmann in the previous section can be applied to obtain the
zeta functions of diagonal surfaces by noting that in the projective case
\[
\Npqks = \frac{q^{ks } -1}{q^k -1} + \eta^{s+1}  q^{k(s-1)/2}B(d,s+1), \; \eta = (-1)^{k+1}.
\]
and hence
\[
\begin{array}{rl}
\ln \left( Z_s (t) \right) = & \sum_{k=1}^{\infty} \Npqks t^k /k\\
      = & \sum_{k=1}^{\infty}  \{ 1 + q^k + \cdots + q^{k(s-1)} + \eta^{s+1}  B(d,s+1) q^{k(s-1)/2} \} t^k /k
\end{array}
\]
If $(s+1)$ is even then $\eta^{s+1}  = 1$ and 
\[
     \ln \left( Z_s (t) \right) =  \sum_{j=0}^{s-1} \{ -\ln (1- q^{j}t)\} -  B(d,s+1) \ln (1-q^{(s-1)/2}t) 
\]
while if $(s+1)$ is odd then  $\eta^{s+1} = (-1)^{(k+1)(s+1)} = (-1)^{k+1}$ and 
\[
     \ln \left( Z_s (t) \right) =  \sum_{j=0}^{s-1} \{ -\ln (1- q^{j}t)\} +  B(d,s+1) \ln (1+q^{(s-1)/2}t) .
\]
Thus the zeta function for diagonal surfaces is given by:
\begin{thm}
\label{thm:diag}
\[
Z_s (t) = \left\{
\begin{array}{ll}
{(1+ q^{(s-1)/2} t)^{B(d,s+1)}}/{\prod_{i=0}^{s-1} (1-q^i t)} , & s ~~ \text{even} \\
{1}/{(1-q^{(s-1)/2} t)^{B(d,s+1)} \prod_{i=0}^{s-1} (1-q^i t)} , & s~~ \text{odd} .
\end{array}
\right.
\]
\end{thm}

The theorem makes explicit the more general Theorem (\ref{thm:weil}) for the 
case that $q=p^{2r}$ and $d \mid (p^r +1)$.

The zeta function for the case $d = p^r +1 = q^{1/2} +1$ i.e. the curve over the base field is Hermitian,
and is given by substituting these parameters in the above expression.

The expressions of the Lemmas (\ref{lemma:1}) to Lemma (\ref{lemma:3}) are sufficiently 
complex that it is an interesting exercise to briefly verify them with the expressions
of Theorem (\ref{thm:wolf}). Only Lemma (\ref{lemma:1}) is considered - Lemma (\ref{lemma:2})
can be dealt with in the same manner. Subtracting $1$ from each side of the equation in
Lemma (\ref{lemma:1}) and dividing by $(q^k - 1)$ gives for the projective 
variety of dimension $s$,
\[
\Npqks  = N_{k,(s-1) }^{\prime} + \naks
\]
and it follows immediately that 
\[
Z_s (t) = Z_{s-1} (t) \cdot F(s,t)
\]
where
\[
F(s,t) = \exp \left( \sum_{k=1}^\infty \naks t^k /k \right)
\]
and the zeta functions for dimensions $s$ and $(s-1)$ are given in the previous theorem
and:
\begin{equation}
\label{eqn:ratio}
\frac{Z_s (t) }{Z_{s-1} (t)}
= \left\{
\begin{array}{rl}
(1+q^{(s-1)/2} t)^{B(d,s+1)} \cdot (1-q^(s-2)/2 t)^{B(d,s)}  / (1-q^{(s-1)} t), &  \text{$s$ even}\\
1/ (1-q^{(s-1)} t) \cdot (1-q^(s-1)/2 t)^{B(d,s+1)}  \cdot  (1+q^{(s-2)/2} t)^{B(d,s)}  , &  \text{$s$ odd}\\
\end{array}
\right.
\end{equation}
To show the right hand side of this equation is $F(s,t)$ notice first that
\[
B(d,s+1) + B(d,s) = (d-1)^s .
\]
and recall $\eta = (-1)^{k+1}$. Then:

\begin{align*}
F(s,t) &= \sum_{k=1}^\infty \naks t^k /k\\ & = \sum_{k=1}^\infty \left\{ q^{k(s-1)} 
+ \eta^{s+1} q^{k(s-2)/2} \big[ (d-1)^s q^{k/2}  - (\eta +q^{k/2} )B(d,s) \big] \right\} t^k /k\\
&  =  \sum_{k=1}^\infty \left\{ q^{k(s-1)} 
+ \eta^{s+1} q^{k(s-2)/2} \big[ q^{k/2} B(d,s+1) - \eta B(d,s) \big] \right\} t^k /k 
\end{align*}
The three terms of the sum are evaluated as follows:

\begin{align*}
\sum_{k=1}^\infty q^{k(s-1)} t^k /k & = \sum_{k=1}^\infty (q^{s-1} t)^k /k = - \ln (1 - q^{s-1} t ) \\
\sum_{k=1}^\infty (-1)^{(s+1)(k+1)} q^{k(s-1)/2} B(d,s+1) t^k /k & = (-1)^{s+1} B(d,s+1) 
    \sum_{k=1}^\infty \big[ (-1)^{s+1} q^{(s-1)/2} t \big]^k /k  \\
& = \left\{
\begin{array}{rl}
B(d,s+1) \big[ - \ln (1 - q^{(s-1)/2} t) \big] ,  &  \text{$s$ odd} \\
B(d,s+1) \big[ + \ln (1 + q^{(s-1)/2} t) \big] , &  \text{$s$ even}
\end{array}
\right.
\end{align*}

and 


\begin{align*}
- \sum_{k=1}^\infty \eta^{s+2} q^{k(s-2)/2} B(d,s) t^k /k 
&= -B(d,s) \sum_{k=1}^\infty (-1)^{s(k+1)} \big( q^{(s-2)/2} t \big)^k /k \\
&=
\left\{
\begin{array}{rl}
- B(d,s) \ln (1 + q^{(s-2)/2} t) , & \quad \text{$s$ odd} \\
+ B(d,s) \ln (1 - q^{(s-2)/2} t ) ,  & \quad \text{$s$ even}
\end{array}
\right.
\end{align*}

Putting the three terms together gives:
\begin{align*}
F(s,t) =
\left\{
\begin{array}{rl}
1 / \big[ (1-q^{(s-1)} t) (1-q^{(s-1)/2} t)^{B(d,s+1)} (1-q^{(s-2)/2} t) \big] ,  & \quad \text{$s$ odd} \\
\big[ (1+ q^{(s-1)/2} t)^{B(d,s+1)} (1 - q^{(s-2)/2} t)^{B(d,s)} \big] / (1-q^{(s-1)} t) , & \quad  \text{$s$ even}
\end{array}
\right.
\end{align*}

Using the form of the zeta functions $Z_s (t)$ and $Z_{s-1} (t)$ of Theorem (\ref{thm:diag})
shows that this is precisely the $Z_s (t) / Z_{s-1} (t)$ given in Equation (\ref{eqn:ratio}).

Recall that the assumption that $d \mid p^r +1$ is a requirement of the results of \cite{wolf}
that is not present in the statement of Theorem(\ref{thm:weil}).

\section{Structure of Hermitian solutions set and coding}
\label{sec:coding}

Results of the previous section are interpreted here for the case of 
Hermitian surfaces which we restate as the set of projective solutions of the
equation
\begin{equation}
\label{eqn:herm}
x_0^{p^r +1} + x_1^{p^r +1} + \cdots + x_s^{p^r +1} = 0,
\end{equation}
over the finite field ${\F_q}$ where $q=p^{2r}$, where $p, ~q,~ r$
are fixed for this section. It will be referred to as the $s$-th dimensional
Hermitian equation. The enumeration
of the size of these varietes is well known \cite{bose,bose1}:
\begin{equation}
\label{eqn:bose}
( p^{r(s-1)} - (-1)^{(s+1)})(p^{rs} - (-1)^s ) / (q-1) = N_{1,s}^{\prime} .
\end{equation}
It is easy to establish the equality of this with
\[
N_{1,s}^{\prime} = \frac{(N_{0,1,s+1} -1)}{(q-1)} = h_s = q^{s-1} + q^{s-2} + \cdots + 1 + q^{(s-1)/2} B(p^r +1, s+1) 
\]
of the previous section. While the Theorem (2.1) was established in \cite{wolf},
using character theory on the finite field, the Hermitian case is far simpler than the 
general diagonal case.  Notice that for an arbitrary $\xi \in {\Fq},~ {\xi}^{p^r +1} \in {\F}_{p^r}^*$. 
In other words the set of $n$-th roots of unity in $\Fq$ are closed under addition,
where $n = (q-1)/(p^r +1) = p^r -1$ and are all the nonzero elements of
${\Fq}^*$.  It follows that the quantity $N_{2,1,s}$ of the previous section is 
identically zero for all $s$.

Consider the set of solutions to the $(s+1)$ dimensional
Hermitian equation. These include The set of solutions to the $s$
dimensional case with a zero added in the $(s+1)$-st position. In addition,
any projective $(s+1)$-tuple that is not a solution to the $s$-dimensional case
can be extended to $(p^r +1)$ solutions to a projective $(s+2)$-tuple that is a solution
to the $(s+1)$ dimensional surface. This implies the size of the $(s+1)$ dimensional 
Hermitian surface is:
\begin{equation}
\label{eqn:rec}
h_{s+1} = (p^r +1)(\pi_s - h_s) + h_s .
\end{equation}
which is essentially Lemma (2.2) with $k=1$, that is this recursion easily leads to the
expression Equation (\ref{eqn:bose}). The three approaches above 
are equivalent. The last approach will be used in the remainder of the section.

The recursion equations of the previous section have an interesting
implication on the recursive structure of the sets of solutions of 
the Hermitian equations. Let $\mathbb{P}^s=\Pqs, ~ (q = p^{2r})$
be the set of projective $(s+1)$-tuples over $\Fq$ and $\mid \mathbb{P}^s \mid = \pi_s$
which for conformity we denote as $p_s$.

To discuss this recursive aspect of the Hermitian solutions sets,
let $P_s$ denote the $(s+1) \times p_s$ 
matrix with columns formed (in some order) from the elements of $\mathbb{P}^s$. 
Similarly denote by $\cHs \subset \mathbb{P}^s$ as the set of 
$(s+1)$-tuples of the $s$-th order Hermitian variety, $h_s = \mid \cHs \mid$ 
and $H_s$ the $(s+1) \times h_s$ matrix of these $(s+1)$-tuples 
over $\Fq$. Further, denote by
$\cAs$ the difference $\mathbb{P}^s \backslash \cHs$ and let  $a_s = p_s - h_s$
and $A_s$ the $(s+1) \times a_s$ matrix of the corresponding $(s+1)$-tuples.

Each $(s+1)$-tuple in $\cAs$ does not satisfy the $s$ dimensional Hermitian equation
Equation (\ref{eqn:herm}). Suppose $(a_0 , a_1 , \cdots , a_s ) \in \cAs$ then
\[
a_0^{p^r +1} + a_1^{p^r +1} + \cdots + a_s^{p^r +1} = b \in {\F}_{p^r}^* ,
\]
i.e. since each of the $(p^r +1)$-th powers of elements in $\Fq$ is in $\F_{p^r}^*$ 
so is their sum (and this is a major difference between the Hermitian and general 
diagonal case where the sum may or may not be in $U_n$). 
Thus, in this case, each such $(s+1)$-tuple in $\cAs$ 
can be extended to $(p^r +1)$ $(s+2)$-tuples
that are solutions to the $(s+1)$ dimensional case i.e.
the equation $x_{s+1}^{p^r +1} = - b \in {\F}_{p^r}$ will have $(p^r +1)$ 
solutions, say $b_1 , b_1 \eta , b_1 \eta^2 , \dots , b_1 \eta^{p^r}$ where $b_1$
is chosen arbitrarily (but fixed) from the set and $\eta$ is a $(p^r +1)$-th root of unity.
Thus $(a_0 , a_1 , \cdots , a_s ) \in \cAs$ yields $(p^r +1)$ solutions to the
$(s+1)$ dimensional case of the form 
$(a_0 , a_1 , \cdots , a_s , b_1^j ) $ of $\cA_{s+1}$.

Let ${\bf b} = (b_1 , b_2 , \dots , b_{a_s} )$ and ${\bf b}_i = \eta^i {\bf b}, ~ i=0,1, \dots, p^r$
where the $b_i$ are representatives of the solutions.
The discussion implies the form of the solutions of to the $(s+1)$-dimensional 
case, (i.e. the elements of $\cH_{s+1}$) in terms of 
the $s$-dimensional case and the corresponding matrices $H_s$ and $H_{s+1}$, and 
reflecting the recursion of Equation (\ref{eqn:rec}), by appropriate permutations 
of the columns, can be viewed as:
\vspace{.2in}

\setlength{\unitlength}{0.25mm}
\begin{center}
\begin{picture}(400,150)
\label{pic:one}
\put(0,0){\line(1,0){400}}
\put(0,0){\line(0,1){150}}
\put(0,150){\line(1,0){400}}
\put(400,0){\line(0,1){150}}
\put(0,130){\line(1,0){400}}
\put(300,0){\line(0,1){150}}
\put(40,0){\line(0,1){150}}
\put(80,0){\line(0,1){150}}
\put(260,0){\line(0,1){150}}
\put(20,60){\makebox(0,0){$A_s$}}
\put(60,60){\makebox(0,0){$A_s$}}
\put(280,60){\makebox(0,0){$A_s$}}
\put(170,60){\makebox(0,0){$\cdots$}}
\put(350,60){\makebox(0,0){$H_s$}}
\put(20,140){\makebox(0,0){${\bf b}$}}
\put(60,140){\makebox(0,0){${\eta \bf b}$}}
\put(280,140){\makebox(0,0){$\eta^{p^r}{\bf b}$}}
\put(420,10){\makebox(0,0){$x_0$}}
\put(420,20){\makebox(0,0){$x_1$}}
\put(420,60){\makebox(0,0){$\vdots$}}
\put(420,125){\makebox(0,0){$x_s$}}
\put(420,140){\makebox(0,0){$x_{s+1}$}}
\put(310,140){\makebox(0,0){$0$}}
\put(320,140){\makebox(0,0){$0$}}
\put(170,140){\makebox(0,0){$\cdots$}}
\put(390,140){\makebox(0,0){$0$}}
\put(350,140){\makebox(0,0){$\cdots$}}
\put(-60,70){\makebox(0,0){$H_{s+1} = $}}
\end{picture}
\vspace{.1in}

{\sc Figure 1}
\end{center}

The above observations have an implication for the structure of the 
evaluation code derived from Hermitian surfaces.
An {\it evaluation code} \cite{tsfa,stic} for a variety $X$ is given by the following map:
\[
\begin{array}{rcl}
\Fq [ x_0 , \dots , x_m ]_h^{\nu} & \longrightarrow & {\Fq}^{n} \\
f & \mapsto & c(f) = (f(P_1 ), f(P_2 ) , \cdots , f(P_n ))
\end{array}
\]
where $P_1 , \dots , P_n$ are a subset of the rational points of the variety
and $f$ a homogeneous polynomial on the $(m+1)$ variables of degree $h$.
When the set of points is the full set of rational points in $\mathbb{P}^s$,
the set of rows generated by monomials of degree $\leq h$ acting on the
$(s +1) \times p_s$ matrix, similar to the Hermitian one above,
form the generator matrix of a code, called the $s$-th order projective geometry
code over $\Fq$. The code is  denoted by $C_P (h,s,q)$ and its parameters
of dimension and minimum distance are well known \cite{lach,lach1,sore1,sore,sore2}. 

Similarly, the code generated and when the full set of points
on the $s$ dimensional Hermitian variety, is denoted by 
$C_H (h,s,q)$ and as noted this is a punctured subcode of 
$C_P (h,s,q)$. Considerably less is known about the Hermitian codes. 

One of the original motivations for this work was to determine 
the recursive structure of the point sets, as above,
and exploit this to determine properties of the Hermitian codes i.e.
it had been hoped to determine properties of the code $C_H (h,s+1,q)$ 
in terms of those of $C_H (h,s,q)$ via the structure of the above matrix.
This seems a difficult task and progress has been limited.
Nonetheless the structure of the matrix is of independent interest in this regard.

It is noted \cite{litt} that for the case of $h=1$ the code $C_H (1,s,q)$
has only two weights,
\[
p^r(2s-1) + (-1)^{s-1} p^{s-1} \quad \text{and} \quad p^{r(2s-1)} .
\] 
This result is directly observable from the above matrix Equation \ref{pic:one}.

A considerable effort has been made (\cite{aubr,bart,cher,edou,edou1,edou2,edou3,hall,sbou}) 
to determine properties of the {\it quadric} Hermitian codes ($h=2$) and 
it has been a challenging problem to determine information on the code.
While some information on this case can be gleaned from the form of the
matrix, one needs information on weights in the projective code
and how they combine with weights of words in the Hermitian code for 
the case of $h=2$ for a given monomial and the situation is more complex.
A new approach to the problem would be of interest.

 \section{Comments}

Some aspects of the structure of the solution sets of multivariate 
diagonal equations over a finite field have been considered and the 
zeta functions of such an equation found explicitly, (which was already implicit
in the work of Wolfmann \cite{wolf}). 

It is noted that Hermitian surfaces, 
where the degree of the equation depends on the size of the 
field of definition, does not have a zeta function. As a matter of interest
define the Hermitian equation
\[
x_0^{p^{kr +1}} + x_1^{p^{kr +1}} + \cdots + x_s^{p^{kr} +1} = 0 .
\]
and seek solutions over ${\F}_{q^k} , ~ q=p^{2r} .$ Thus the degree of the Hermitian equation increases with the 
size of the extension field. One could define a 'zeta like' function by examining the enumeration
as $k$ increases. Let $N^{\prime}_{k,s}$ denote the number of projective solutions 
to this equation over ${\F}_{q^k}$ and note that:
\[
\begin{array}{rl}
N^{\prime}_{k,s} & = (q^{ks} -1)/{(q-1)} + q^{k(s-1)/2} B(p^{rk} +1,s+1)\\
        & = \left\{    
\begin{array}{rl}
        \pi_{q^k , s-1} + p^{rks} \{ \sum_{j=0}^{a-1} p^{2jrk +1} - \sum_{j=0}^{a-1} p^{2rkj} \}         & s = 2a ,  \\
        \pi_{q^k , s-1} + p^{rks} \{ \sum_{j=0}^{a} p^{2jrk} - \sum_{j=0}^{a-1} p^{2rkj+1} \}         & s = 2a+1 , 
\end{array}
\right.
\end{array}
\]
Noting that 
\[
\sum_{k=1}^{\infty} p^{jrsk} t^k /k = - \ln (1 - p^{jrs} t)
\]
define the polynomials
\[
P(t) = \prod_{i=1}^{ks-1} (1-q^i t), \; P_1 (t) = \prod_{i= 0}^{a-1} (1-p^{r(s+2j)}t)  .
\]
With this definition the resulting `zeta function' is:
\[
Z(t) = \exp \{ \sum_{k=1}^{\infty} N^{\prime}_{k,s} t^k /k \}
   = \left\{
\begin{array}{rl}
\frac{1}{P(t) \cdot P_1 (t)^{p-1}} , & s=2a \\
\hspace{1in} \\
\frac{P_1 (t)^{(p-1)}}{(1- p^{r (s+2a)}t) \cdot P(t) } , & s=2a+1 .
\end{array}
\right.
\]
This is to be compared to the Theorem (\ref{thm:diag}) with $d=p^r +1$ where the 
diagonal equation degree is constant.

\end{document}